\pdfoutput=1
\documentclass[11pt,twoside,a4paper,cmspaper,final,collab]{cms-tdr}
\def\svnVersion{ad00f39}\def\svnDate{2024/09/03}\def\cmsCernNoTag{CERN-EP-2023-300}\def\cmsCernDate{\today}\def\cmsMessage{Published in Physics Letters B as \href{http://dx.doi.org/10.1016/j.physletb.2024.138964}{\doi{10.1016/j.physletb.2024.138964}.}}
\begin{document}\cmsNoteHeader{HIG-21-017}

\newlength\cmsTabSkip\setlength{\cmsTabSkip}{1ex}

\newcommand{\massj}{\ensuremath{m^{AK8}}\xspace}
\newcommand{\ptj}{\ensuremath{\pt^{AK8}}\xspace}
\newcommand{\pth}{\ensuremath{\pt^{\PH}}\xspace}
\newcommand{\ptm}{\ensuremath{\pt^{\PGm}}\xspace}
\newcommand{\pte}{\ensuremath{\pt^{\Pe}}\xspace}
\newcommand{\ptt}{\ensuremath{\pt^{\PGt_{\mathrm{h}}}}\xspace}
\newcommand{\ptz}{\ensuremath{\pt^{\PZ}}\xspace}
\newcommand{\Imu}{\ensuremath{I_{rel}^{\PGm}}\xspace}
\newcommand{\Ie}{\ensuremath{I_{rel}^{\Pe}}\xspace}
\newcommand{\Ime}{\ensuremath{I_{rel}^{\PGm,\Pe}}\xspace}
\newcommand{\ljpt}{\ensuremath{\pt^{\mathrm{j}_1}}\xspace}
\newlength\cmsFigWidth
\ifthenelse{\boolean{cms@external}}{\setlength\cmsFigWidth{0.85\columnwidth}}{\setlength\cmsFigWidth{0.4\textwidth}}
\ifthenelse{\boolean{cms@external}}{\providecommand{\cmsLeft}{upper\xspace}}{\providecommand{\cmsLeft}{left\xspace}}
\ifthenelse{\boolean{cms@external}}{\providecommand{\cmsRight}{lower\xspace}}{\providecommand{\cmsRight}{right\xspace}}
\ifthenelse{\boolean{cms@external}}{\providecommand{\cmsTable}[1]{#1}}{\providecommand{\cmsTable}[1]{\resizebox{\textwidth}{!}{#1}}}
\newcommand{\emu}{\ensuremath{\Pe\PGm}\xspace}
\newcommand{\mue}{\ensuremath{\PGm\Pe}\xspace}
\newcommand{\tautau}{\ensuremath{\tauh\tauh}\xspace}
\newcommand{\mutau}{\ensuremath{\PGm\tauh}\xspace}
\newcommand{\etau}{\ensuremath{\Pe\tauh}\xspace}

\cmsNoteHeader{HIG-21-017}
\title{Measurement of the production cross section of a Higgs boson with large transverse momentum in its decays to a pair of \texorpdfstring{$\PGt$}{tau} leptons in proton-proton collisions at \texorpdfstring{$\sqrt{s} = 13\TeV$}{sqrt(s) = 13 TeV}}

\date{\today}
\abstract{
A measurement of the production cross section of a Higgs boson with transverse momentum greater than~250\GeV is presented where the Higgs boson decays to a pair of \PGt leptons. It is based on proton-proton collision data collected by the CMS experiment at the CERN LHC at a center-of-mass energy of 13\TeV. The data sample corresponds to an integrated luminosity of 138\fbinv. Because of the large transverse momentum of the Higgs boson the \PGt leptons from its decays are boosted and produced spatially close, with their decay products overlapping. Therefore, a dedicated algorithm was developed to reconstruct and identify them. The observed (expected) significance of the measured signal with respect to the standard model background-only hypothesis is 3.5 (2.2) standard deviations. The product of the production cross section and branching fraction is measured to be $1.64^{+0.68}_{-0.54}$ times the standard model expectation. The fiducial differential production cross section is also measured as functions of the Higgs boson and leading jet transverse momenta. This measurement extends the probed large-transverse-momentum region in the $\PGt\PGt$ final state beyond 600\GeV.}
\hypersetup{
pdfauthor={CMS Collaboration},
pdftitle={Measurement of the production cross section of a Higgs boson with large transverse momentum in its decays to a pair of \texorpdfstring{$\PGt$}{tau} leptons in proton-proton collisions at \texorpdfstring{$\sqrt{s}=13\TeV$}{sqrt(s) = 13 TeV}},
pdfsubject={Boosted H tautau},
pdfkeywords={CMS, Higgs boson, tau lepton}} 

\maketitle 

\section{Introduction}
\label{sec:introduction}

Since the discovery of a particle compatible with the standard model (SM) Higgs boson ($\PH$)~\cite{Aad:2012tfa, Chatrchyan:2012xdj, Chatrchyan:2013lba}, many of its properties have been scrutinized, without evidence of deviations from the SM expectations~\cite{ATLAS:2022vkf,CMS:2022dwd}. 
Measuring the production of the Higgs boson with large transverse momentum (\pt) tests the existence of physics beyond 
the SM (BSM) in the scalar sector, to which inclusive measurements could be insensitive~\cite{Battaglia:2021nys}. Such measurements can shed light on the structure of the interactions of the Higgs boson with massive particles. In particular, the cross section of the gluon-gluon production mode of the Higgs boson (ggH) is sensitive at large \pt to the structure of the effective ggH coupling~\cite{Grojean:2013nya,Grazzini:2016paz}.

The ATLAS and CMS Collaborations have reported measurements of the differential production cross section of the Higgs boson as a function of 
its transverse momentum \pth~\cite{ATLAS:2023jdk,ATLAS:2023hyd,ATLAS:2022yrq,ATLAS:2022qef,ATLAS:2022fnp, ATLAS:2021tbi,ATLAS:2020wny, CMS:2024ddc,CMS:2023gjz, CMS:2022wpo, CMS:2021gxc,CMS:2020zge, CMS:2018gwt}. While the measurement precision at low $\pth$ is driven by the $\PH\to\PGg\PGg$, $\PH\to\PZ\PZ\to4\ell$, and $\PH\to\PW\PW\to 2\ell2\nu$ decay modes ($\ell$= $\Pe$ or $\PGm$), the sensitivity at larger $\pth$ is improved by including the $\PH\to\PGt\PGt$  and $\PH\to\cPqb\cPqb$ decay modes because of their larger branching fractions.
In this Letter, a dedicated measurement of the production cross section of a Higgs boson with large \pt decaying to a pair of $\PGt$ leptons is presented. This measurement extends the probed large-\pt region beyond 600\GeV~\cite{CMS:2021gxc}. Because of the large-\pt of the Higgs boson, the $\PGt$ leptons from its decays are boosted and produced spatially close with their decay products overlapping. A Higgs boson with $\pth\gtrsim250\GeV$ leads to a separation $\Delta R = \sqrt{\smash[b]{(\Delta \eta)^2 + (\Delta \phi)^2}}<0.8$ between the $\PGt$ leptons. To handle overlapping $\PGt$ decay products, a dedicated algorithm to identify the close-by $\PGt$ leptons is employed.
All four main Higgs boson production modes have been included in this measurement and four final states including hadronic and leptonic $\PGt$ decays are considered: $\mutau$, $\etau$, $\tautau$, and $\emu$, where $\tauh$ denotes the visible decay products of a $\PGt$ lepton decaying into hadrons.
The results are extracted using a multiclass neural network (NN) categorizing the events into signal- and background-enriched regions.
The analysis is based on proton-proton ($\Pp\Pp$) collision data collected by the CMS experiment at the CERN LHC at a center-of-mass energy of 13\TeV. The data sample corresponds to an integrated luminosity of 138\fbinv~\cite{CMS-LUM-17-003,CMS-PAS-LUM-17-004,CMS-PAS-LUM-18-002}. 

\section{The CMS detector}
\label{sec:detector}

The central feature of the CMS apparatus is a superconducting solenoid of 6\unit{m} internal diameter, providing a magnetic field of 3.8\unit{T}. Within the solenoid volume are a silicon pixel and strip tracker, a lead tungstate crystal electromagnetic calorimeter, and a brass and scintillator hadron calorimeter, each composed of a barrel and two endcap sections. Forward calorimeters extend the pseudorapidity ($\eta$) coverage provided by the barrel and endcap detectors. Muons are detected in gas-ionization chambers embedded in the steel flux-return yoke outside the solenoid. 
Events of interest are selected using a two-tiered trigger system. The first level, composed of custom hardware processors, uses information from the calorimeters and muon detectors to select events at a rate of around 100\unit{kHz} within a fixed latency of 4\mus~\cite{CMS:2020cmk}. The second level, known as the high-level trigger, consists of a farm of processors running a version of the full event reconstruction software optimized for fast processing, and reduces the event rate to around 1\unit{kHz} before data storage~\cite{CMS:2016ngn}.
A more detailed description of the CMS detector, together with a definition of the
coordinate system used and the relevant kinematic variables, can be found in
Ref.~\cite{Chatrchyan:2008zzk}.

\section{Event simulation}
\label{sec:simulation}

The \POWHEG 2.0~\cite{Nason:2004rx,Frixione:2007vw, Alioli:2010xd, Alioli:2008tz}  generator is used to generate Higgs boson signal samples at next-to-leading order (NLO) in 
the four leading production modes. The distributions of $\pth$
 and jet multiplicity in the gluon fusion production simulation are corrected to
match the predictions of the NNLOPS event generator based upon the $\POWHEG+$MiNLO approach~\cite{Hamilton:2013fea,Hamilton:2015nsa}.
The $\PW+\text{jets}$ and $\PZ/\PGg^*+\text{jets}$ processes are modelled with 
 \MGvATNLO 2.2.2 (2.4.2) generator~\cite{Alwall:2014hca} with the MLM jet matching and merging scheme~\cite{Alwall:2007fs}.
A combination of the \POWHEG 2.0~\cite{Melia:2011tj,Nason:2013ydw} and \MGvATNLO generators is used to model the diboson ($\PW\PW$, $\PW\PZ$, and $\PZ\PZ$) processes at NLO.
Single top quark and \ttbar productions are modeled with the \POWHEG~\cite{Alioli:2009je,Re:2010bp} generator at NLO.  
The NNPDF 3.1~\cite{Ball:2014uwa} parton distribution functions (PDF) are used, 
and all simulated samples are interfaced with \PYTHIA 8.212~\cite{Sjostrand:2014zea} 
to describe parton showering and hadronization. The {CUETP8M1}~\cite{Khachatryan:2015pea} and {CP5}~\cite{CMS:2019csb} tunes 
are used for the 2016 and 2017--2018 samples, respectively.
Additional inelastic $\Pp\Pp$ interactions (pileup) generated by \PYTHIA are overlaid on all simulated events, according to the
luminosity profile of the analyzed data. 
The average pileup ranges from 23 interactions in 2016 to 32 in 2018.
All the generated signal and background samples are processed with the simulation of the CMS detector based on \GEANTfour~\cite{Agostinelli:2002hh}. 
Differences between data and simulation in trigger, particle identification, isolation efficiencies, and in the resolution 
of the jet $\pt$ and missing transverse momentum ($\ptvecmiss$) are corrected by applying scale factors to simulated events.

\section{Event reconstruction and selection}
\label{sec:reconstruction}

The physics objects are reconstructed by the particle-flow 
(PF) algorithm~\cite{Sirunyan:2017ulk}, which combines information from all 
subdetectors to identify individual particles (PF candidates).
The primary vertex is taken to be the vertex corresponding to the hardest scattering in the event, evaluated using tracking information alone, as described in Section 9.4.1 of Ref.~\cite{CMS-TDR-15-02}.
Standard muon and electron reconstruction, identification, and isolation methods are used, and are similar to those 
described in Ref.~\cite{CMS:2021gxc}.
Starting from PF candidates, AK4 and AK8 jets are reconstructed using the anti-$\kt$ \FASTJET algorithm with a distance parameter of 0.4 and 0.8, 
respectively~\cite{Cacciari:2008gp, Cacciari:2011ma}.
Jet energy corrections are applied to both simulation and data so that the average measured energy of jets becomes identical to that of simulated particle-level jets, to account for residual differences between the jet energy in data and in simulation~\cite{CMS:2016lmd,CMS:2020ebo}, and to reduce pileup effects.

The PF candidates in AK4 jets are used to reconstruct $\tauh$ objects, using the hadrons-plus-strips (HPS) algorithm~\cite{TAU-11-001,Khachatryan:2015dfa}, which combines one or three charged particles with neutral pion candidates within a narrow signal cone, to identify the $\tauh$ decay modes. 
PF candidates inside a cone of size $\Delta R=0.5$, defined as an isolation cone,
around the $\tauh$ candidate are used to determine its isolation.
To distinguish genuine $\tauh$ decays from jets originating from the hadronization of quarks or gluons, a multivariate analysis (MVA) discriminant is used~\cite{CMS:2018jrd}, called ``MVA isolation" in the following. It combines information about the energy carried by PF candidates within the isolation cone with lifetime information.
The lifetime information includes the impact parameter of the leading track of the tau candidate and its significance. In the case of 3-prong tau leptons, the secondary vertex is fitted using the tau tracks and a flight path length, i.e. a signed distance between the primary vertex and secondary vertex, and its significance are used in addition to the impact parameter.  
Electrons and muons reconstructed as $\tauh$ candidates are rejected with dedicated discriminants~\cite{CMS:2018jrd}. 

If a $\PGt$ lepton pair originates from a particle with high \pt, the decay products of the 
$\PGt$ leptons are emitted close to each other, forming a single jet. The separation of the $\PGt$ leptons is approximately equal to $\frac{2m_{H}}{\pth}$.
The presence of two close-by $\PGt$ leptons spoils the isolation requirement, leading to a drop in efficiency. 
To recover this loss, 
a modified version of the HPS algorithm, 
relying on jet substructure techniques, has been developed~\cite{CMS:2018jrd}. Particles are first clustered 
in a large-radius jet with the Cambridge--Aachen algorithm~\cite{Wobisch:1998wt} with a distance parameter of 0.8, and the jet is required to have $\pt>170\GeV$. 
The final step of the jet clustering is reversed to find two subjets consistent with $\PGt$ leptons. 
The subjets are then processed with the HPS algorithm to identify the $\tauh$ decay modes.
A crucial step in the boosted $\tauh$ reconstruction algorithm is the requirement that the overlapping $\PGt$ leptons do not share any common PF candidate. This is, by construction, satisfied by the two subjets seeding the $\PGt$ lepton reconstruction algorithm. Thus, 
the isolation of a given $\PGt$ is computed after removal of the PF candidates that belong to the other $\PGt$ lepton.
Since electrons and muons give rise to subjets, the method also excludes these leptons from the list of neighboring tau lepton isolation candidates in the \etau and \mutau  final states.
The final step is to compute discriminants to reject jets, electrons, and muons mis-reconstructed as $\tauh$ candidates. 

\begin{figure*}[htbp]
  \centering
    \includegraphics[width=0.49\textwidth]{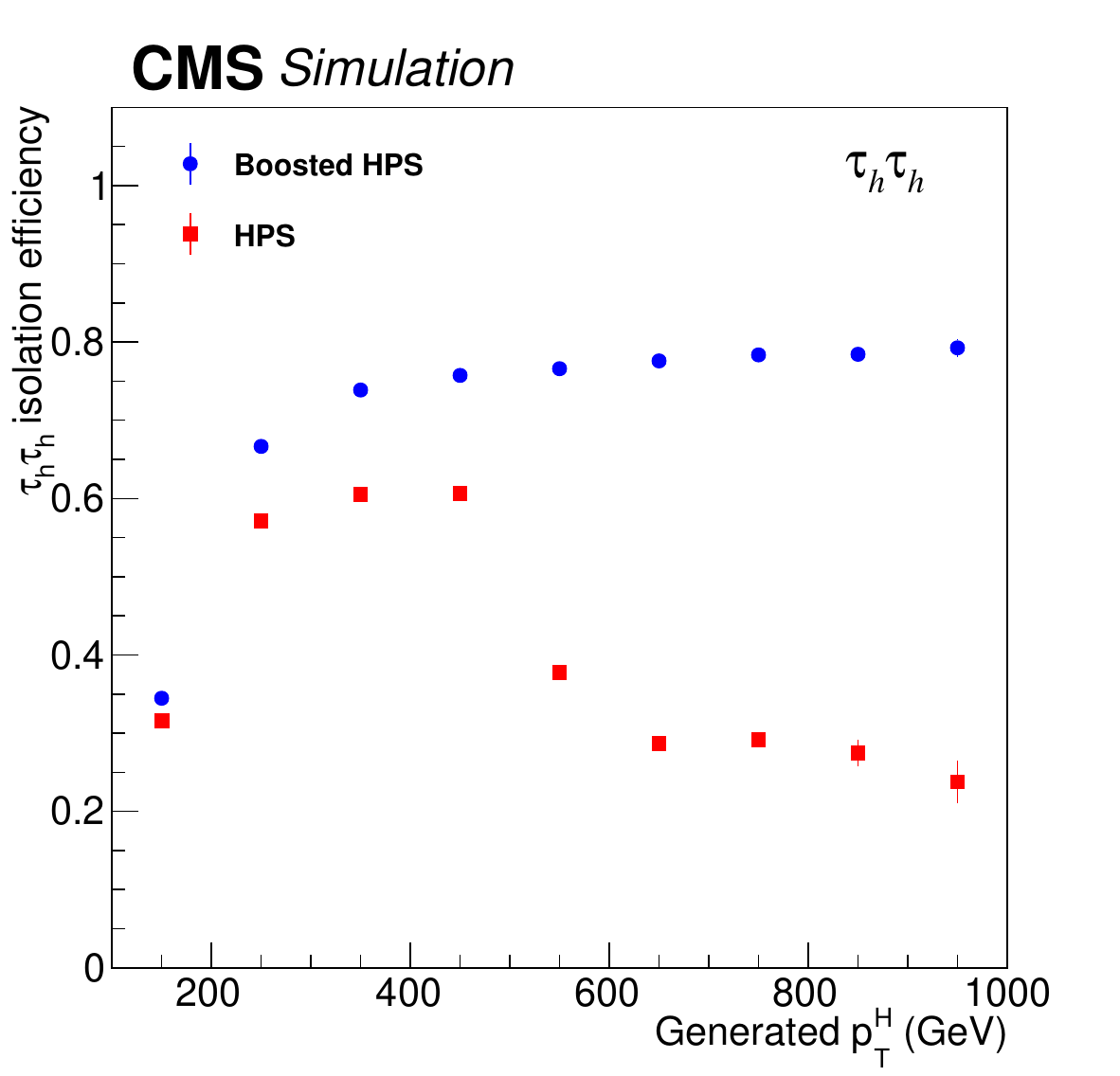}
    \includegraphics[width=0.49\textwidth]{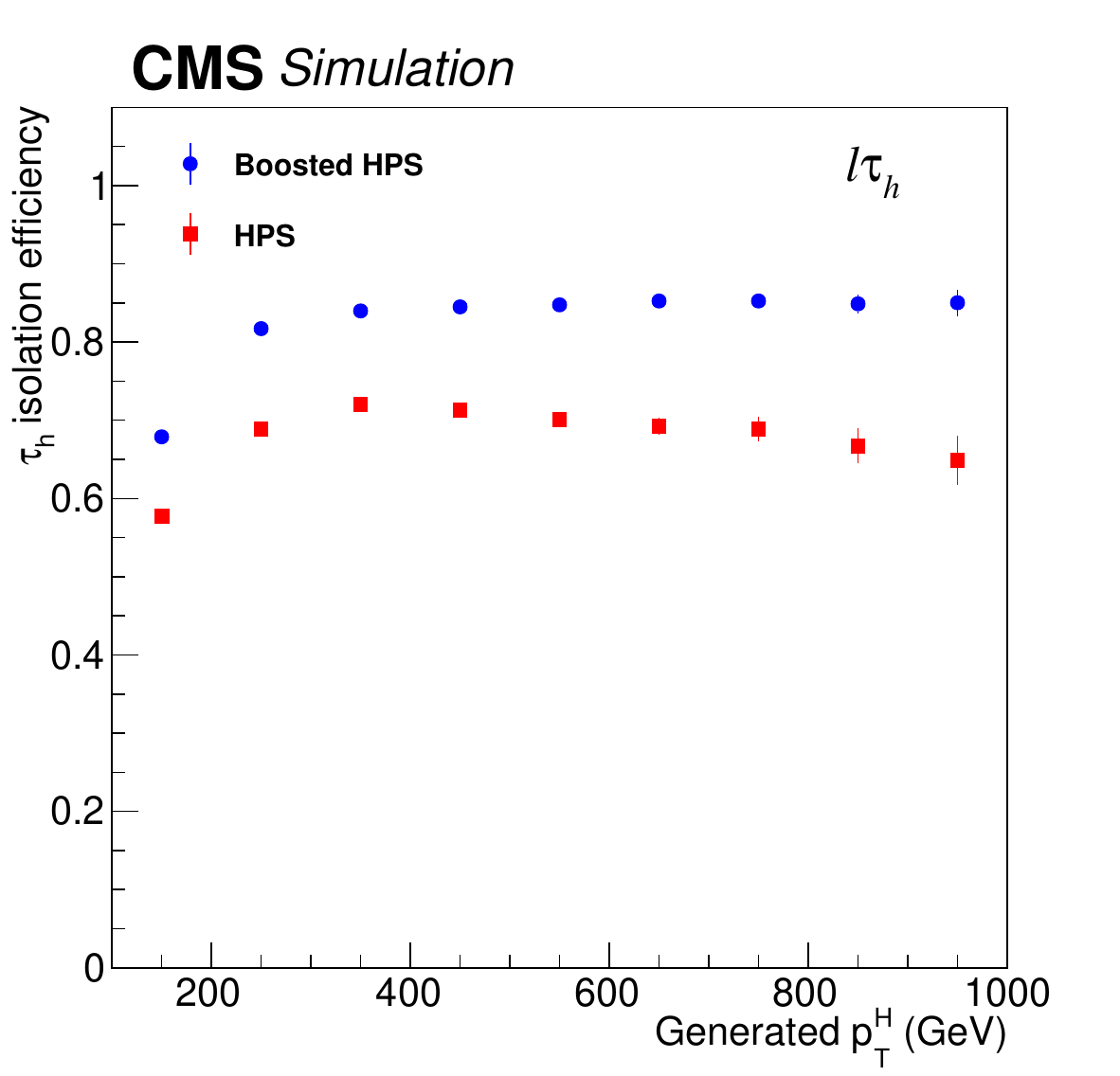}
                  
	  \caption{Comparison of the isolation efficiencies for the standard and boosted HPS algorithms in the \tautau (left) and $\ell\tauh$ (right) final states.}
    \label{fig:tauLepComparsion}
\end{figure*}

Figure~\ref{fig:tauLepComparsion} compares the $\tauh\tauh$ and $\tauh$ isolation efficiencies for the standard and boosted HPS algorithms as a function of \pth, in the \tautau (left) and $\ell\tauh$ (right) final states. As $\pth$ increases and the $\PGt$ leptons get spatially closer, the HPS $\PGt$ lepton isolation 
efficiency decreases while the boosted $\PGt$ lepton efficiency reaches a plateau and remains constant. This effect is more pronounced 
in the \tautau final state because of the presence of a pair of $\tauh$ candidates. 
 A drop in the isolation efficiency of HPS $\tauh$ candidate around 450\GeV occurs when the distance between the $\tauh$ candidate becomes equal to the size of the $\tauh$ isolation cone of 0.5.

The $\tauh$ candidates must have $\pt>30\GeV$, $\abs{\eta}<2.3$, and pass an MVA isolation requirement that corresponds to a typical identification efficiency of 60\% in $\PZ/\PGg^*\to\PGt\PGt$ events for a 3\% misidentification rate of jets.
 The loose working point of the discriminator against muons (electrons) is used, suppressing muons (electrons) by more than 90\% and maintaining the $\tauh$ identification efficiency above 95\%.
The  $\ptvecmiss$ is computed as the negative vector sum of the transverse momenta of all the PF candidates in an event, and its magnitude is denoted as \ptmiss~\cite{CMS:2019ctu}. The \ptvecmiss is modified to account for corrections to the energy scale of the reconstructed jets in the event. 
Another correction accounts for the effect of additional pileup interactions~\cite{CMS:2020ebo}. 
Two variables to quantify the hadronic activity in the event are used. The first one is the total 
visible hadronic activity \HT, which is defined as the scalar sum of transverse momenta of jets with $\pt>30\GeV$ and $\abs{\eta}<3$. 
The second variable is the momentum imbalance \htvecmiss and is defined as the negative vector sum of the transverse momenta of jets with $\pt>30\GeV$ and $\abs{\eta}<5$. Its magnitude is denoted as \mht. 

Collision events were collected using single-lepton (muon or electron) triggers in the $\mutau$, $\etau$, and $\emu$ channels, while a combination of triggers relying on $\ptmiss$, jet \pt, $\HT$, and $\mht$ was used in the $\tauh\tauh$ channel. In the $\mutau$ ($\etau$) channel, events passing single-lepton triggers with \pt threshold of 50 (115)\GeV and no isolation requirement must have an offline lepton with $\pt>52\ (120)\GeV$, while events passing single isolated lepton triggers must have an offline lepton with \pt between 28 
(38) and 52 (120)\GeV and passing isolation criteria, as well as $\ptmiss>30\GeV$. Events in the \emu channel must pass any of the single-lepton triggers, with or without isolation, depending on the offline leading lepton \pt. The lepton not associated with the trigger must have \pt larger than 10\GeV and pass loose identification and isolation criteria. 

In the $\tautau$ channel, events triggered in 2016 (2017-2018) on the basis of an AK8 jet with $\pt>360$ (400)\GeV and mass, using the soft drop algorithm~\cite{Larkoski:2014wba}, greater than 30\GeV are required to have an AK8 jet reconstructed offline with $\pt>450\GeV$ and mass above 30\GeV. Events that do not pass the AK8 jet trigger are then checked against a combination of $\ptmiss$-based trigger criteria.Events that pass an online selection based on $\HT>300$ (500)\GeV, $\ptmiss > 110$ (100)\GeV, and $\mht> 0$ (100)\GeV, are required to have offline $\HT> 400$ (700)\GeV and  $\ptmiss >180\GeV$ ($\ptmiss + \mht>280\GeV$). The trigger selection efficiency for events passing the offline requirements described here for the $\tautau$ channel is about 90\%  across all three years. The trigger requirements and kinematic selection criteria are summarized in Table~\ref{tab:EventSelection}.

\begin{table*}[htbp]
\centering
	  \topcaption{Selection requirements for the four $\PGt\PGt$ final states. The relative isolation variable, \Ime, for muons and electrons is evaluated as the scalar sum of the \pt of the reconstructed particles in a cone around the lepton track relative to the \pt of the lepton.  In the \tautau channel, the trigger requirement is defined by a combination of trigger candidates above a given threshold, indicated inside parentheses in \GeV.  The thresholds for the offline  selection are driven by the trigger requirements. Except for the \Ime, the quantities are in units of \GeV.}
\cmsTable{
\begin{tabular}{lclclc|}
Final state &  Trigger & Selection criteria \\
\hline
\mutau                       & $\PGm(50)$  & \ptm $>$  52 , \ptt $>$  30  \\
\mutau                       & Iso $\PGm(27)$  &  \ptm $>$  28 , \ptt $>$  30 , $\Imu< 0.3$, $\ptmiss > 30$ \\
\etau                       & e($115$)  & \pte $>$  120 , \ptt $>$  30  \\
\etau                       & Iso e$(35)$  &  \pte $>$  38 , \ptt $>$  30 , $\Ie< 0.3$, $\ptmiss > 30$ \\
\mue                       & $\PGm(50)$  & \ptm $>$  52 , \pte $>$  10  \\
\mue                       & Iso $\PGm(27)$ &  \ptm $>$  28 , \pte $>$  10 , $\Ime< 0.3$, $\ptmiss > 30$ \\
\emu                       & e($115$)  & \pte $>$  120 , \ptm $>$  10  \\
\emu                       & Iso e$(35)$  &  \pte $>$  38 , \ptm $>$  10 , $\Ime< 0.3$, $\ptmiss > 30$ \\
\tautau(2016)                       &  \ptj(360) \& \massj(30) & 			 $\ptj > 450, \massj > 30$     \\
\tautau (2016)                      & \HT(300) \& \ptmiss(110)   &  			 $\HT >  400  , \ptmiss >180$    \\
\tautau(2017--2018)              &  \ptj(400) \& \massj(30)  &  	  	$\ptj > 450, \massj > 30$     \\
\tautau (2017--2018)             & \HT(500) \& \ptmiss(100) \& \mht(100)   &   $\HT >  700 , (\ptmiss + \mht)>280$    \\
\end{tabular}
}
\label{tab:EventSelection}
\end{table*}

To reconstruct the Higgs boson candidates, the two identified objects ($\Pe$, $\PGm$, or $\tauh$ candidates) are required to have opposite charges. None 
of the jets in the event must be identified 
as originating from a bottom quark, using the medium (tight) working point of the 
 \textsc{DeepCSVv2} algorithm~\cite{Sirunyan:2017ezt} in the \mutau and \etau (\emu) final states. 
The quantity $\HT$ must be greater than 200\GeV and the di-$\PGt$ invariant mass, estimated from the visible momenta of the $\PGt$ lepton decay products 
and $\ptvecmiss$, using a maximum likelihood technique~\cite{Bianchini:2016yrt}, must exceed 50\GeV.
The reconstructed $\pth$ is defined as the
magnitude of the vectorial sum of the \pt of the $\PGt$ leptons visible decay products and $\ptvecmiss$, and 
must be larger than 250\GeV. Finally, to reduce the background with leptonically decaying \PW bosons, the transverse mass of the lepton and $\ptvecmiss$ in the $\mutau$ and $\etau$ channels, 
and of the dilepton and $\ptvecmiss$ in the $\emu$ channel, must be less than 80\GeV.
To ensure there is no overlap across different final states, events are vetoed in the presence of additional loosely identified and isolated lepton.

The dominant background in all four final states is the Drell--Yan process, which constitutes an irreducible background. This is followed by events with at least one jet being mis-identified as a $\tauh$, muon, electron, and quantum chromodynamics (QCD) multijet events with nonprompt leptons. The \ttbar and diboson processes constitute the remaining background.

\section{Background estimation}
\label{sec:background}

The Drell--Yan process is modelled using simulated samples.
 To correct the effect of missing higher order 
corrections in simulation, a scale factor that depends on the $\PZ/\PGg^*$ boson mass and \pt is measured in a control region (CR) containing  
a pair of muons, and applied to the simulation in the signal region (SR). 
The \ttbar background is estimated from simulation. Simulated \ttbar events are weighted to reproduce the observed distribution of the top quark \pt and to account for any possible missing higher order corrections~\cite{CMS:2012hkm,CMS:2015rld}.
The background arising from the misidentification of leptons or $\tauh$ candidates, called "mis-ID background", is estimated from data. 
It is mostly comprised of $\PW+\text{jets}$ and QCD multijet events, 
where at least one jet is misidentified as an electron, muon, 
or $\tauh$ candidate. 
In events with at least one $\tauh$ candidate, a misidentification rate method is used to estimate the contribution 
of these two processes, collectively, in the SR. 
The probability for a jet that has passed a very loose isolation to pass the nominal isolation requirement of a $\tauh$ candidate, called the misidentification rate, is 
estimated in a region dominated by the mis-ID background, 
obtained by requiring one identified and isolated muon, and one jet separated from the muon by $\Delta R>0.8$. 
This sample has a negligible signal component, while the small contribution from the other backgrounds is estimated from simulation and subtracted from data. 
The misidentification rate, $f$, is parameterized as a function of the \pt of the $\tauh$ candidate.
To estimate the mis-ID background in the SR, an adjusted misidentification rate, $\frac{f}{1-f}$, is applied as a weight to events in
the application region in data. The application region definition is identical to that of SR with the exception that the $\tauh$ candidate
fails the nominal MVA isolation. 
The denominator accounts for the fraction $(1-f)$ of mis-ID background events failing the isolation. The small contribution from the signal and other backgrounds is estimated from simulation and subtracted from data before the reweighting is applied.

To estimate the QCD multijet contribution in the $\emu$ channel, a sample is built from events that pass the signal 
selection criteria except that the requirement on the charge of two leptons is inverted.
The small remaining contribution of the signal and other backgrounds is subtracted from data on the basis of simulation.
The contribution of the QCD multijet background in the SR is derived by applying an opposite-sign to same-sign scale factor, 
measured in a region where the lepton isolation requirements are inverted.

\section{Analysis strategy}
\label{sec:strategy}

A multiclass NN is used to construct a discriminant 
that separates the signal from the two dominant background processes, \ie Drell--Yan and mis-ID backgrounds, following closely the 
method described in Ref.~\cite{Collaboration:2022mlq}. 
The NN is constructed with a feed-forward architecture containing two hidden layers 
and three output nodes.  The number of
nodes per layer is kept to a minimum to reduce the complexity of the neural network without compromising its performance.
The value of each output node represents the probability for an event to belong to a given class.
 Ten variables are used as input to the NN:
 the \pt of the two leptons and leading jet, $\ptmiss$, the mass and \pt of the Higgs boson reconstructed 
from the $\PGt$ candidates and $\ptvecmiss$, $\mathrm{S_{T}}$ (defined as the scalar sum of the \pt of all leptons and jets with $\pt>30\GeV$ and $\abs{\eta}<3$ in the event), 
the transverse mass reconstructed from the lepton \pt and $\ptvecmiss$, the value of the MVA isolation
of the closest boosted $\tauh$ candidate to the muon (electron) in the $\mutau$ and $\emu$ ($\etau$ and $\emu$) channels, 
and the di-$\PGt$ mass. The NN is trained for the combination of the three data-taking years and separately for each final state.

\begin{figure*}[htbp]
\centering
\includegraphics[width=0.49\textwidth]{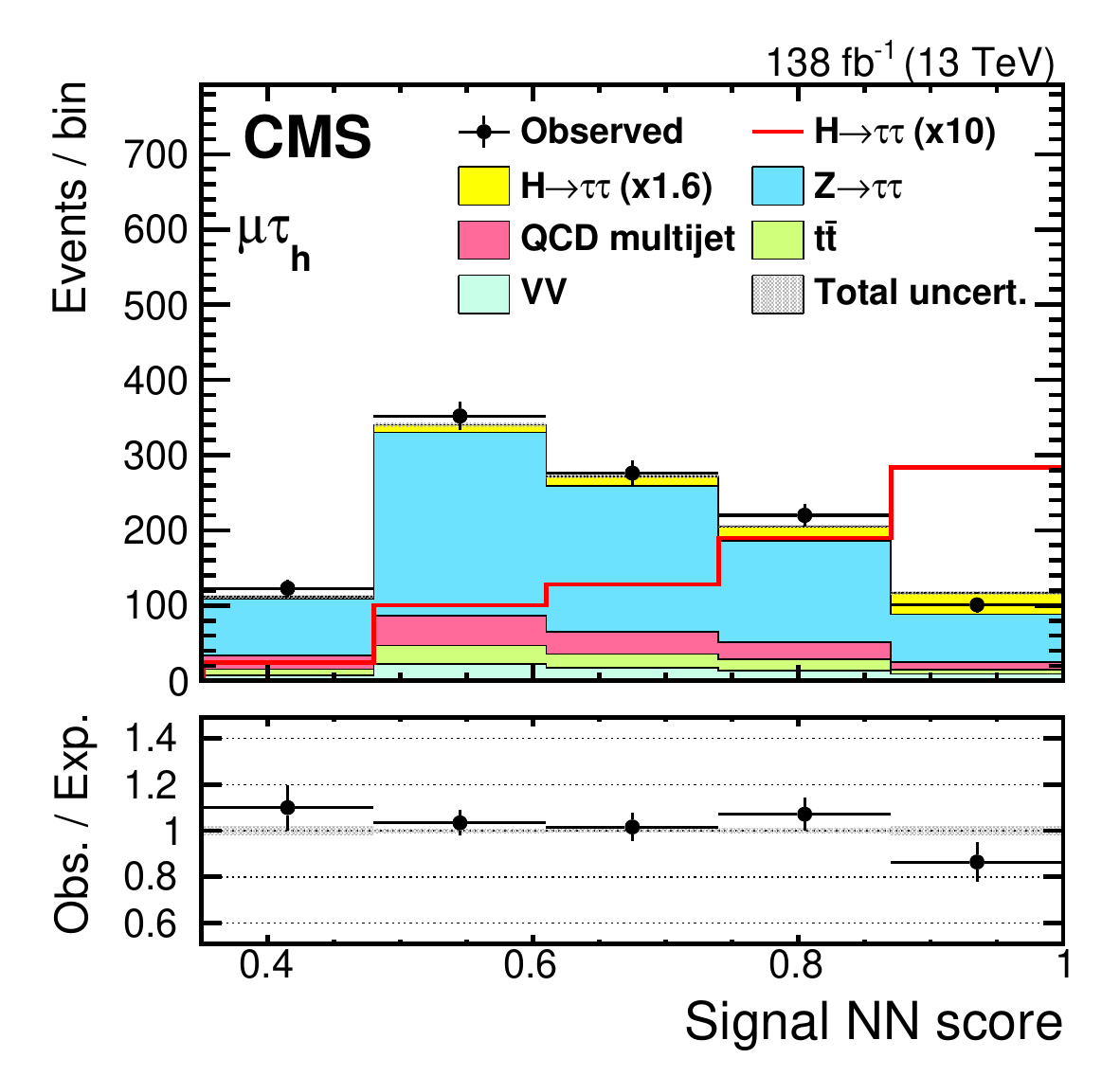}
\includegraphics[width=0.49\textwidth]{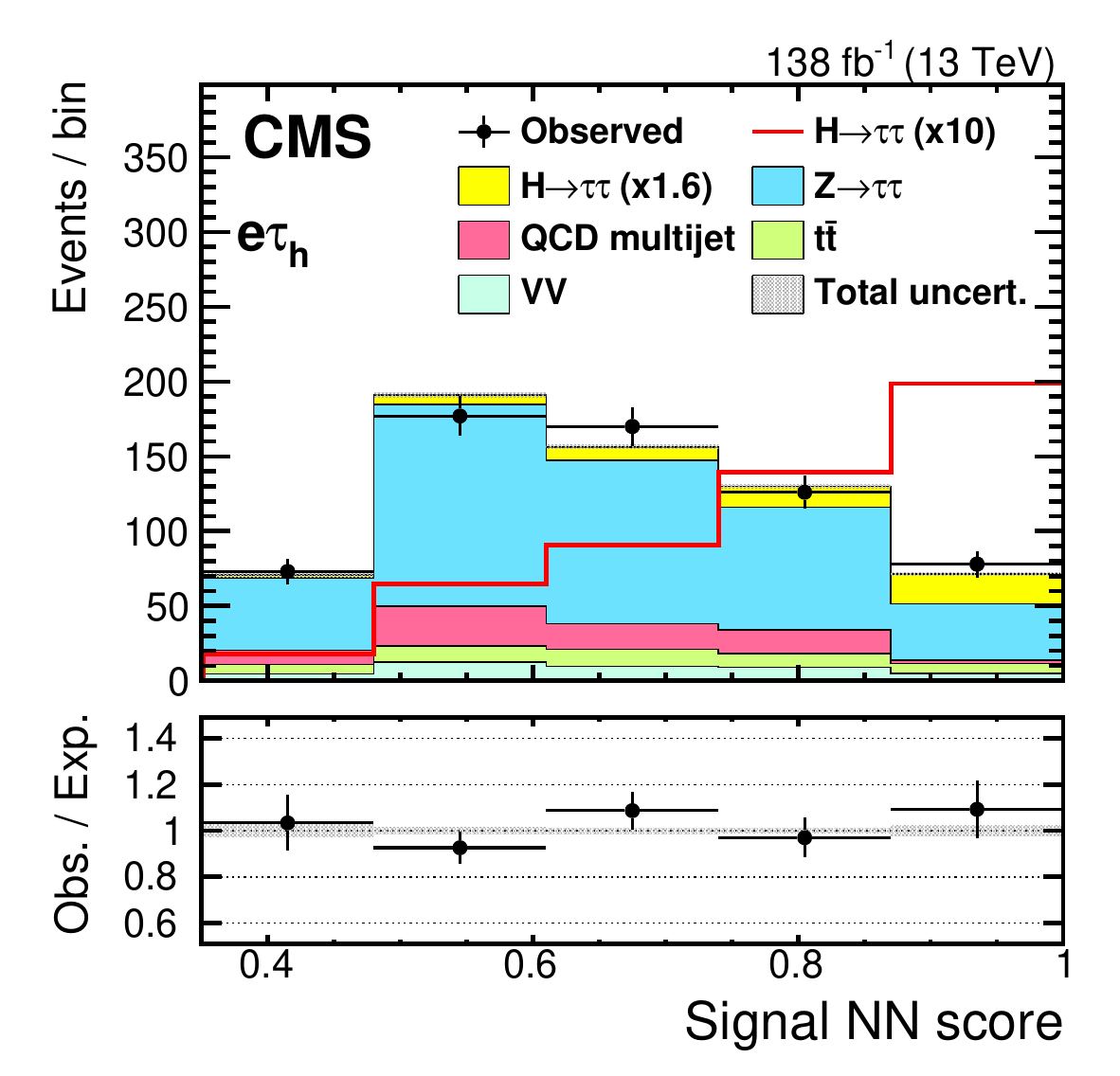}\\
\includegraphics[width=0.49\textwidth]{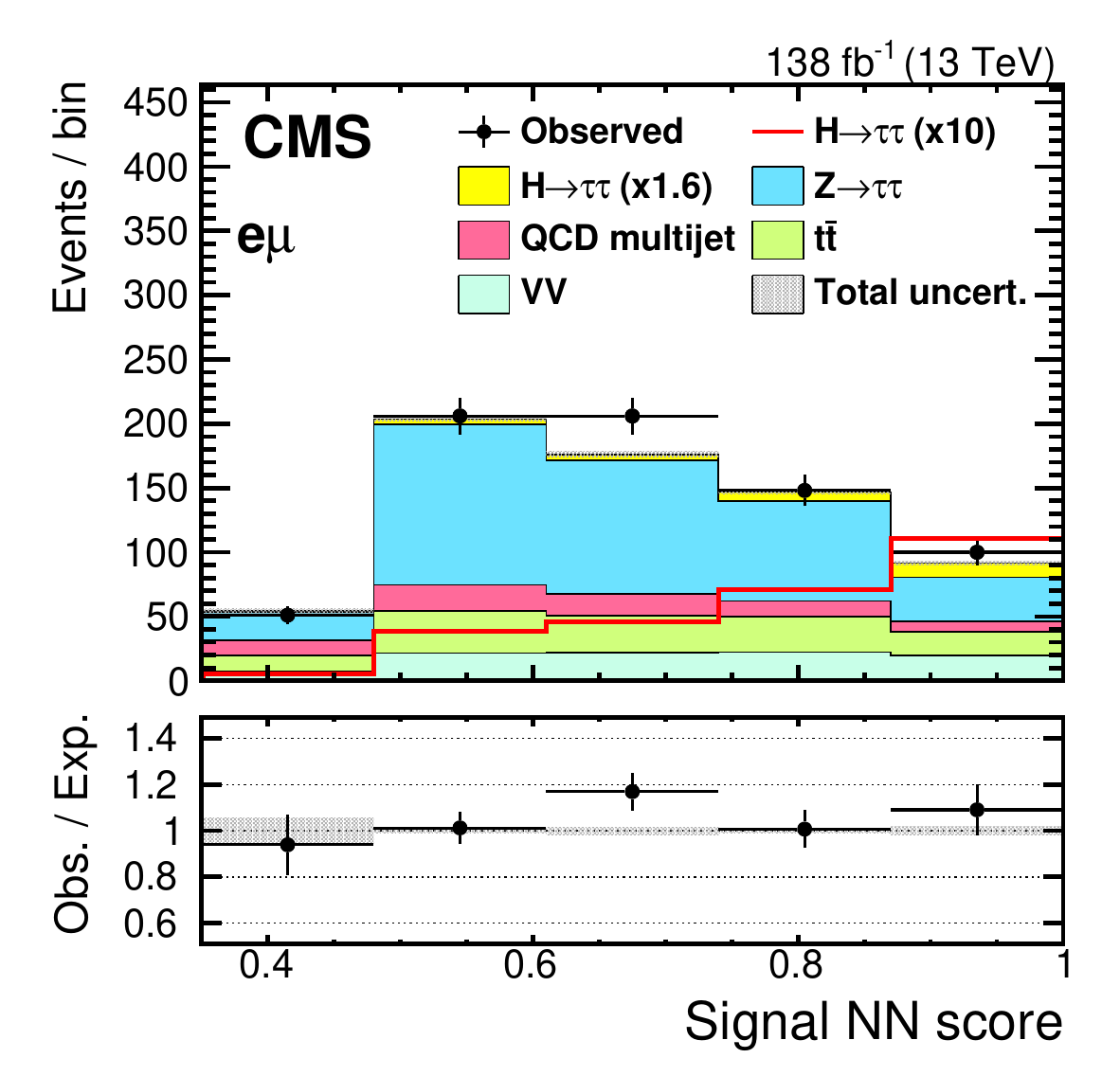}
\includegraphics[width=0.49\textwidth]{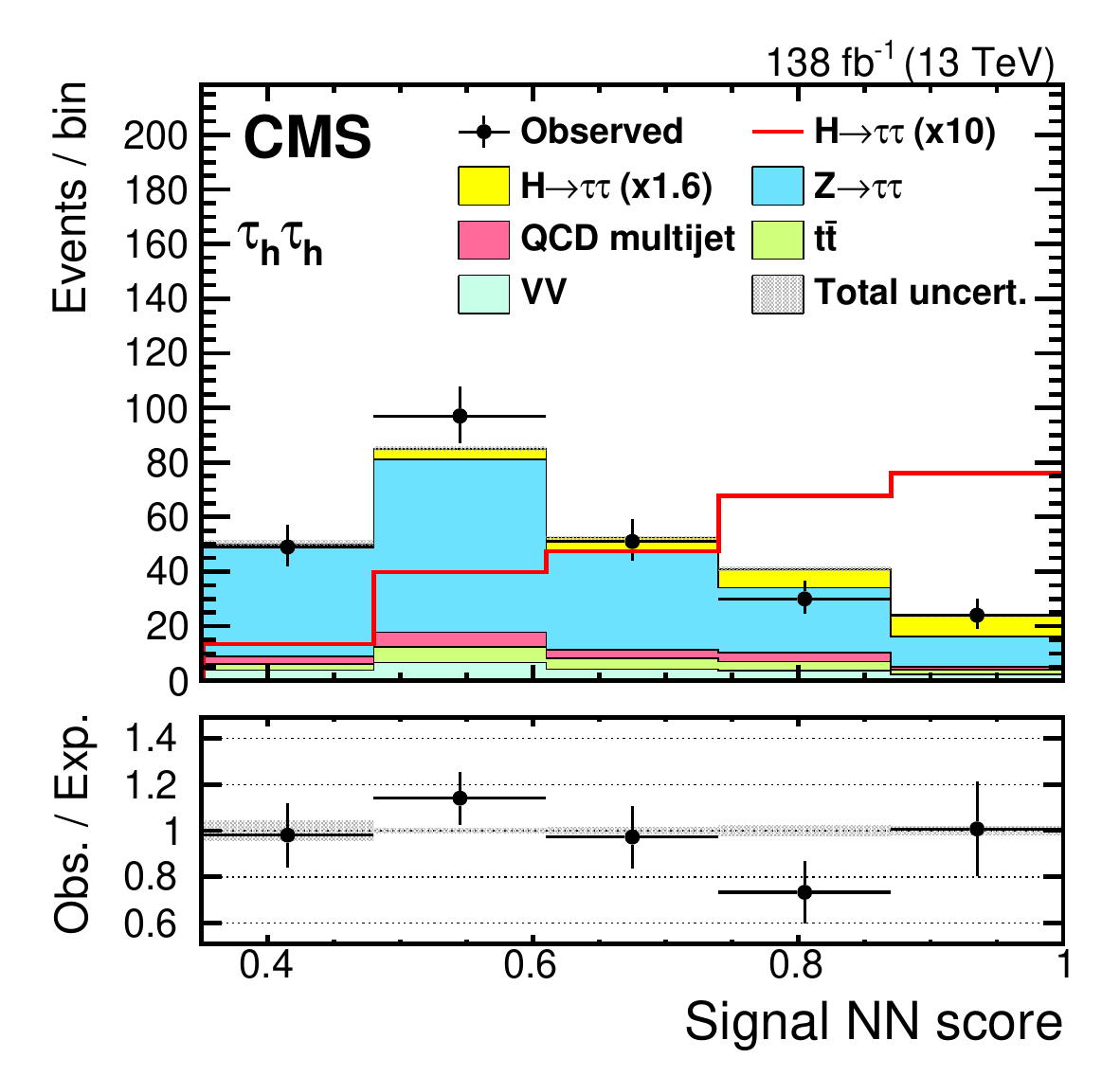}
\caption[dummy text]{Observed and expected NN distributions in the signal-enriched region, after combining all four $\pth$ bins, in the $\mutau$ (upper left), $\etau$ (upper right), $\emu$ (lower left), and $\tautau$ (lower right) channels. The signal and background distributions are the result of a simultaneous binned maximum likelihood fit to all three output NN distributions, including all individual \pt bins and data-taking years. The bottom panel shows the ratio of the number of events observed in data to that of the expected background.}
\label{fig:NN2016}
\end{figure*}

Based on the maximum of the three output values, by construction larger than 1/3, events are sorted into one signal-enriched region and two background-dominated regions.
The signal-enriched region is split into four bins depending on $\pth$, with lower bin boundaries of 250, 350, 450, and 600\GeV. To establish the presence of a signal a simultaneous binned maximum likelihood fit is performed to all three NN output distributions.
The combined NN distributions in the signal-enriched region for all four $\pth$ bins and three data-taking years are shown in Fig.~\ref{fig:NN2016}.

\section{Systematic uncertainties}
\label{sec:systematics}

Two types of systematic uncertainties are considered: normalization uncertainties in the event yields and uncertainties that affect both the shape and yield of the distributions.
The uncertainties in the muon and electron trigger efficiencies, identification, and isolation, estimated with a tag-and-probe method~\cite{Khachatryan:2015dfa}, range from 1 to 5\%, depending on the \pt and $\eta$ of the lepton.

The uncertainty in the $\HT$ and $\ptvecmiss$ trigger efficiencies in the $\tautau$ channel is 10\%, mostly arsing from the difference in trigger efficiency in data and simulation.
The uncertainty in the efficiency of selecting boosted $\tauh$ candidates is parametrized as a function of \pth. 
To evaluate this uncertainty, a maximum likelihood fit is performed using the \PZ boson \pt distributions reconstructed using the dilepton four momenta and $\ptmiss$. The fit is performed in the Drell--Yan-dominated region, which is by definition orthogonal to the signal-enriched region.
A large prefit uncertainty ($U_{prefit}$) is assigned to the $\tauh$ identification efficiency, uncorrelated between final states, data-taking years, 
and bins of $\ptz$. This is equivalent to letting this uncertainty freely float in the fit.
After the fit, the magnitude of the postfit $\tauh$ identification efficiency uncertainty ($U_{postfit}$) is compared to $\abs{U_{prefit} - U_{postfit}}$.
The largest value in each final state, data-taking years and $\ptz$ range is taken as the uncertainty in the $\tauh$ identification efficiency. It typically ranges between 10 and 20\% except for the very last $\ptz$ bin in the \tautau final state
where it reaches 50\%.

 The uncertainty in the integrated luminosity amounts to 1.2--2.5\%, depending on the year~\cite{CMS-LUM-17-003,CMS-PAS-LUM-17-004,CMS-PAS-LUM-18-002}.
A 2\% normalization uncertainty is considered for the muon/electron energy scale, while a 3\% uncertainty in the 
$\tauh$ energy scale is treated as a shape uncertainty that modifies final observable templates and propagated to the di-$\PGt$ mass and $\ptvecmiss$.
Uncertainties associated with the jet energy scale
and $\ptmiss$ are also considered~\cite{CMS-PAS-JME-16-003}.

Uncertainties of 2.0, 4.2, 5.0, and 5.0\% are used for the predicted cross sections of the Drell--Yan, \ttbar, single top quark, and diboson productions, respectively~\cite{Campbell:2011bn, Czakon:2011xx, Gehrmann:2014fva, Melnikov:2006kv}. 
The PDF, renormalization and factorization scales, initial and final state radiation uncertainties are considered for the Drell--Yan and \ttbar processes. They vary from a few percent to about 20\% depending on the final state, process, and data-taking year as they are evaluated with different Pythia tunes.

A 20\% uncertainty is assigned to the QCD multijet background to cover for differences between the observed and predicted background rates in dedicated CRs.
An additional uncertainty related to the probability for a jet to be misidentified as a $\tauh$ candidate, 
as well as an uncertainty
in the normalization of the non-QCD background subtracted from data in the application region, are included and vary 
between 10 and 20\%, depending on the final state.
Uncertainties related to the top quark \pt reweighting in simulated \ttbar events
are evaluated by varying the reweighting parameters between zero
and twice their nominal values~\cite{CMS:2012hkm,CMS:2015rld}.

Theoretical uncertainties for the signal in the gluon fusion production mode are derived from the LHC Higgs Cross Section WG1 scheme~\cite{LHCHiggsCrossSectionWorkingGroup:2016ypw}. 
Statistical uncertainties associated with the limited number of simulated events in the signal-enriched region or observed event yields in the control regions are accounted for as described in Ref.~\cite{BarlowBeeston}.

\section{Results}
\label{sec:results}
 
A binned maximum likelihood fit is performed to all three NN output distributions 
to compute the probability of the compatibility of observed data with the background-only 
hypothesis~\cite{Conway:2011in,CMS-NOTE-2011-005}. 
The inclusion of the two background-dominated control regions help to constrain the normalization and systematic uncertainties of the Drell--Yan and mis-ID backgrounds. 
The systematic uncertainties  are taken as nuisance parameters.
An observed significance of 3.5 standard deviations  is obtained for a $\PH\to\PGt\PGt$ signal with $\pth\geq250\GeV$ by combining all channels and 
data-taking years, to be compared with an expected significance of 2.2 standard deviations. 
The best fit signal strength modifier, defined as the ratio of the inclusive Higgs boson production cross section to the SM expectation, is also 
extracted from a maximum likelihood fit to the same distributions and is found to be $1.64^{+0.68}_{-0.54}$, 
in agreement with the SM prediction.
The sensitivity of this analysis is driven by the $\mutau$ and $\etau$ channels in the low-\pt region, and by the $\tautau$ channel in the high-\pt region. The sensitivity of the $\tautau$ channel increases with the Higgs boson \pt because of the increase in efficiency of the \HT/\ptmiss triggers.

The differential production cross section is measured in a fiducial phase space that is defined per final state.
Requirements on generator-level variables are chosen to be close to the trigger and offline selections.
In the $\mutau$ ($\etau$) channel, the lepton is required to have generated $\pt>28\ (30)\GeV$ and $\abs{\eta}<2.4\ (2.5)$. If the generated \pt 
of the muon (electron) is less than 52 (115)\GeV, a minimum threshold of 30\GeV is applied to the generated $\ptmiss$ as well. 
The generated visible \pt of  $\tauh$ leptons must exceed 30\GeV and their $\abs{\eta}$ must be less than  2.3 in the \mutau, \etau, and \tautau channels.
In the $\tautau$ channel, the generated \pt of the AK8 jet must be greater than 450\GeV or the generated $\HT$ must be greater than 400 (700)\GeV with a generated $\ptmiss$ greater than 180 (140)\GeV for 2016 (2017--2018) samples.
In the $\emu$ channel, two sets of events are selected. The first set includes events with generated muon $\pt > 52\GeV$,
or between 28 and 52\GeV and $\ptmiss$ above 30\GeV. The second set includes events with generated
electron $\pt>115\GeV$, or between 38
and 115\GeV and $\ptmiss$ above 30\GeV. In both sets, the other lepton must have generated \pt above 10\GeV.
In all channels, the $\Delta R$ between two leptons must be between 0.1 and 0.8, the generated $\pth$ must be above 250\GeV, 
and there must be at least one jet with \pt larger than 30\GeV.

\begin{figure}[thbp]
  \centering
    \includegraphics[width=0.45\textwidth]{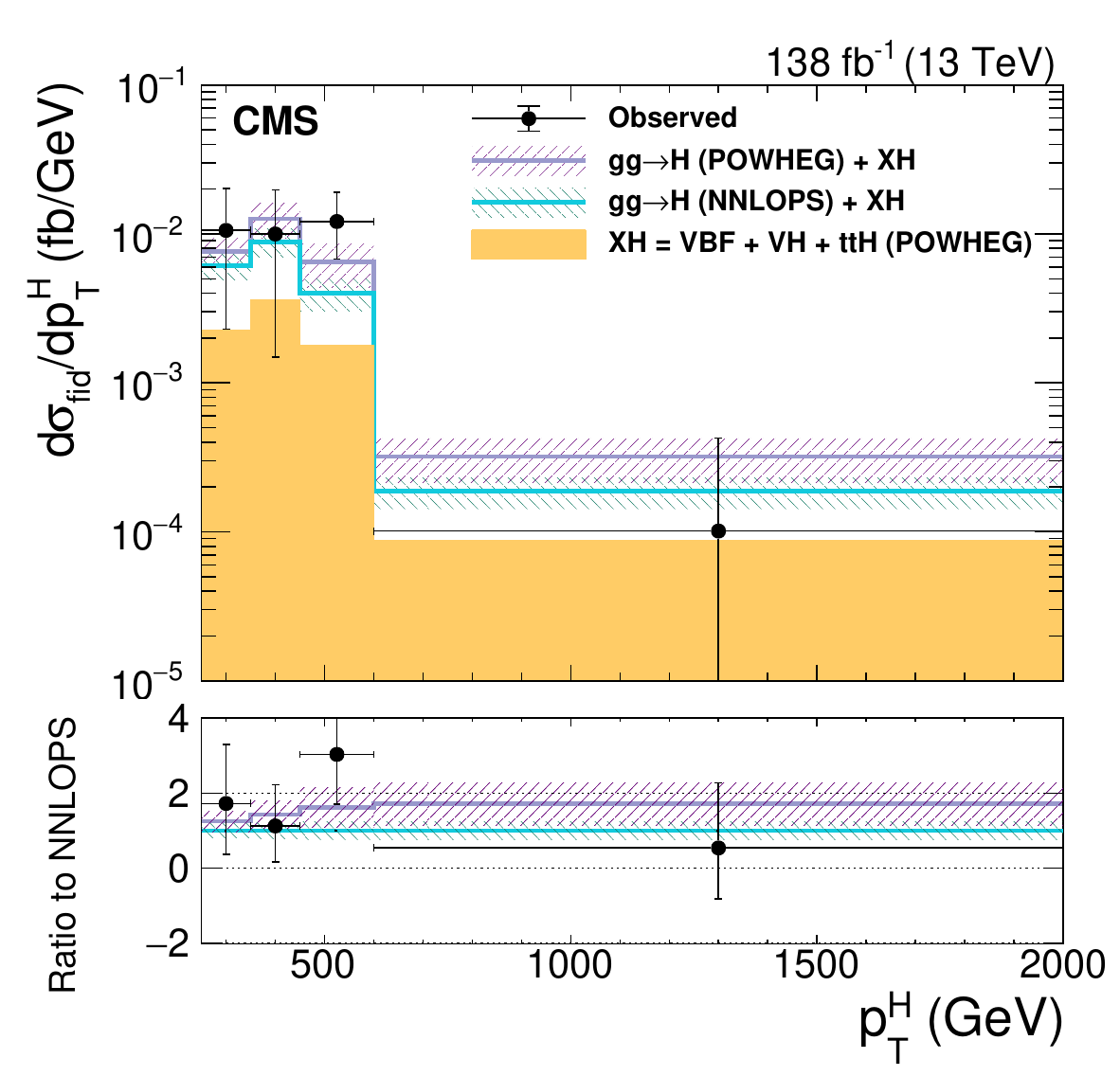}
    \includegraphics[width=0.45\textwidth]{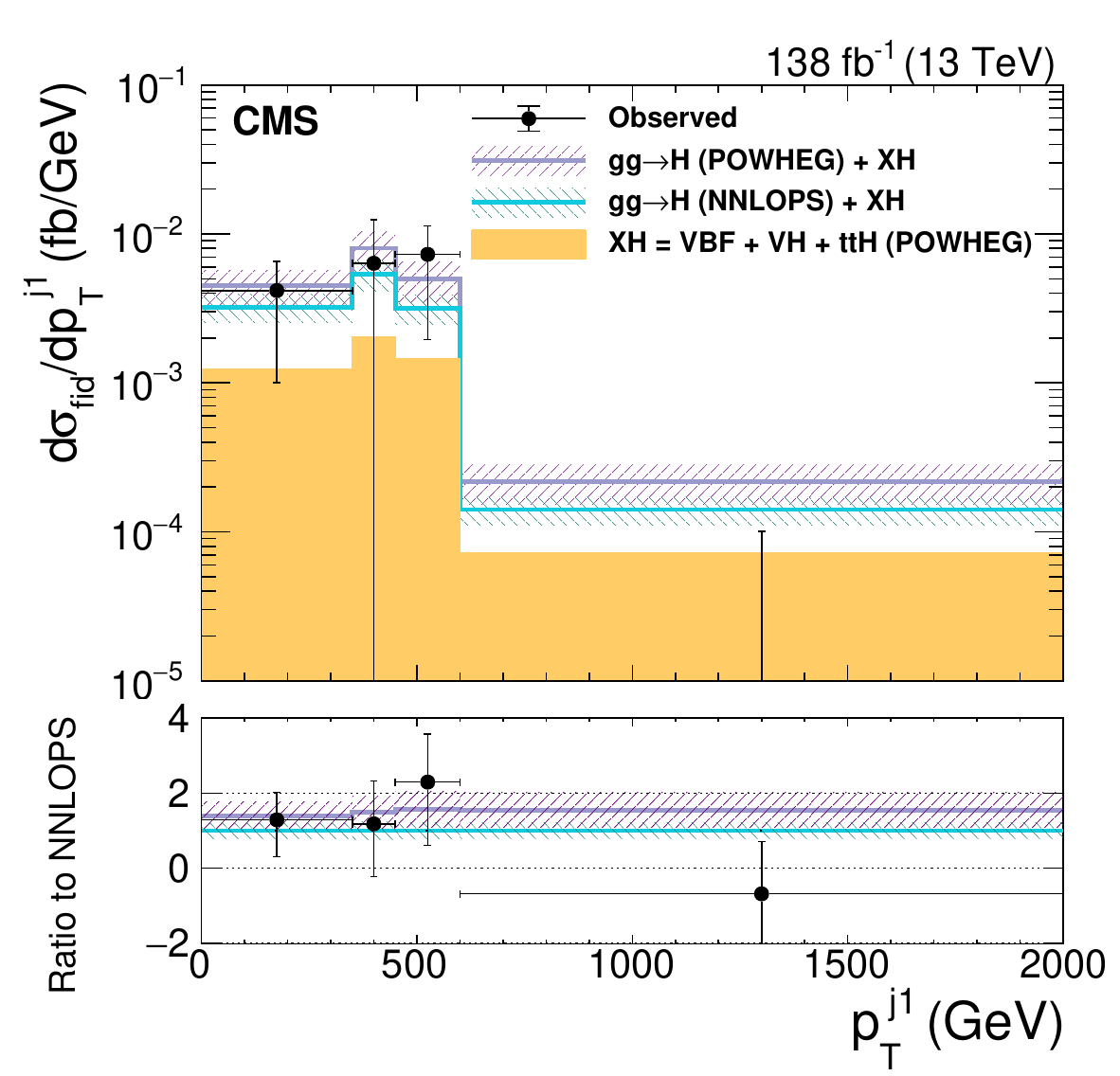}
    \caption{Observed and expected differential fiducial cross sections in bins of $\pth$
(\cmsLeft) and $\ljpt$ (\cmsRight). The last bins include the overflow. The uncertainty bands in the theoretical predictions include uncertainties
from the following sources: PDF, renormalization and factorization scales, underlying event
and parton showering, and the branching fraction $\mathcal{B}(\PH\to\PGt\PGt)$. }
    \label{fig:HiggspT}
\end{figure}

To derive the differential production cross section, the signal is split into four bins with lower bin boundaries of 250, 350, 450, and 600\GeV on the value of the generated (gen) $\pth$ or with lower bin boundaries of 0, 350, 450, and 600\GeV on the value of the gen
leading jet \pt ($\ljpt$). The definitions of the signal- and  background-dominated regions are identical to those used in the inclusive analysis.
The same binning is used at reconstructed- (reco-) level  to categorize events.
The gen- and reco-level observable values are not perfectly aligned 
because of the limited resolution, and events from one gen-level bin can migrate to  
a different reco-level bin. The contributions from the four gen-level bins are left floating independently from each other. 
By performing one simultaneous fit over all reco-level bin histograms of the NN distributions, the signal strength modifiers of the gen-level
observable bins are determined utilizing the full statistical power of the data set. This is equivalent to extracting 
the signal in the reco-level bins and performing an unfolding to gen-level bins. 
The procedure follows the strategy adopted in Ref.~\cite{CMS:2021gxc}.
Fiducial differential production cross sections measured as functions of $\pth$  and $\ljpt$ are shown in Fig.~\ref{fig:HiggspT}. Tabulated results are available in the HepData database~\cite{HEPData}. 
The results are dominated by statistical uncertainties, especially in the highest-\pt bins. 
The next dominant components,  especially in the lowest-\pt bins, are the tau energy scale uncertainty, which affects both the normalization and shape of the signal and background processes. Other important systematic uncertainties include theoretical uncertainties in the signal prediction arising from the infinite top mass assumption and initial and final states radiation.
The ggH production mode contribution in the total signal yield reduces from 63\% in the lowest $\pth$ bin to 53\% in the highest $\pth$ bin, indicating the importance of other production modes at large \pt values. This measurement extends the probed large-\pt region beyond 600\GeV. No significant deviation with respect to the SM predictions is observed and the measurements are compatible with both the \POWHEG and NNLOPS expectations. The fit has a $p$-value with respect to the SM expectation from the NNLOPS prediction of 56 and 74\% for the measurements with respect to $\pth$ and $\ljpt$, respectively. 

The inclusive fiducial production cross section is measured from the $\pth$ distributions used in the differential analysis, by reformulating the parameters of interest such that one modifies the total inclusive fiducial production  cross section. Its best fit value is $3.88^{+1.69}_{-1.35}$ fb, which is consistent with the SM prediction of $2.36\pm0.51$ fb obtained with the NNLOPS generator.  

\section{Summary}
\label{sec:summary}

The measurement of the production cross section of a  Higgs boson with large \pt decaying to a pair of $\PGt$ leptons has been performed using proton-proton collision data collected by the CMS experiment at a center-of-mass energy of 13\TeV, corresponding to an integrated luminosity of 138\fbinv. A dedicated reconstruction algorithm has been used to resolve the overlapping decay products of the two close-by $\PGt$ leptons. The $\PH\to\PGt\PGt$ signal with $\pth>250\GeV$ is established with a significance of 3.5 standard deviations (2.2 expected). The best fit of the product of the observed $\PH\to\PGt\PGt$ signal production cross section and branching fraction is $1.64^{+0.68}_{-0.54}$ times the SM expectation. The fiducial inclusive production cross section has been measured to be $3.88^{+1.69}_{-1.35}$ fb, which is consistent with the SM prediction of $2.36\pm0.51$ fb. The fiducial differential production cross section is also measured as functions of the Higgs boson and leading jet transverse momenta. This measurement extends the probed large-transverse-momentum region beyond 600\GeV. No significant deviation with respect to the SM predictions is observed in the transverse momentum distribution of the Higgs boson with large \pth.

\begin{acknowledgments}
  We congratulate our colleagues in the CERN accelerator departments for the excellent performance of the LHC and thank the technical and administrative staffs at CERN and at other CMS institutes for their contributions to the success of the CMS effort. In addition, we gratefully acknowledge the computing centers and personnel of the Worldwide LHC Computing Grid and other centers for delivering so effectively the computing infrastructure essential to our analyses. Finally, we acknowledge the enduring support for the construction and operation of the LHC, the CMS detector, and the supporting computing infrastructure provided by the following funding agencies: SC (Armenia), BMBWF and FWF (Austria); FNRS and FWO (Belgium); CNPq, CAPES, FAPERJ, FAPERGS, and FAPESP (Brazil); MES and BNSF (Bulgaria); CERN; CAS, MoST, and NSFC (China); MINCIENCIAS (Colombia); MSES and CSF (Croatia); RIF (Cyprus); SENESCYT (Ecuador); ERC PRG, RVTT3 and MoER TK202 (Estonia); Academy of Finland, MEC, and HIP (Finland); CEA and CNRS/IN2P3 (France); SRNSF (Georgia); BMBF, DFG, and HGF (Germany); GSRI (Greece); NKFIH (Hungary); DAE and DST (India); IPM (Iran); SFI (Ireland); INFN (Italy); MSIP and NRF (Republic of Korea); MES (Latvia); LMTLT (Lithuania); MOE and UM (Malaysia); BUAP, CINVESTAV, CONACYT, LNS, SEP, and UASLP-FAI (Mexico); MOS (Montenegro); MBIE (New Zealand); PAEC (Pakistan); MES and NSC (Poland); FCT (Portugal); MESTD (Serbia); MCIN/AEI and PCTI (Spain); MOSTR (Sri Lanka); Swiss Funding Agencies (Switzerland); MST (Taipei); MHESI and NSTDA (Thailand); TUBITAK and TENMAK (Turkey); NASU (Ukraine); STFC (United Kingdom); DOE and NSF (USA).
   
  \hyphenation{Rachada-pisek} Individuals have received support from the Marie-Curie program and the European Research Council and Horizon 2020 Grant, contract Nos.\ 675440, 724704, 752730, 758316, 765710, 824093, 101115353,101002207, and COST Action CA16108 (European Union); the Leventis Foundation; the Alfred P.\ Sloan Foundation; the Alexander von Humboldt Foundation; the Science Committee, project no. 22rl-037 (Armenia); the Belgian Federal Science Policy Office; the Fonds pour la Formation \`a la Recherche dans l'Industrie et dans l'Agriculture (FRIA-Belgium); the Agentschap voor Innovatie door Wetenschap en Technologie (IWT-Belgium); the F.R.S.-FNRS and FWO (Belgium) under the ``Excellence of Science -- EOS" -- be.h project n.\ 30820817; the Beijing Municipal Science \& Technology Commission, No. Z191100007219010 and Fundamental Research Funds for the Central Universities (China); the Ministry of Education, Youth and Sports (MEYS) of the Czech Republic; the Shota Rustaveli National Science Foundation, grant FR-22-985 (Georgia); the Deutsche Forschungsgemeinschaft (DFG), under Germany's Excellence Strategy -- EXC 2121 ``Quantum Universe" -- 390833306, and under project number 400140256 - GRK2497; the Hellenic Foundation for Research and Innovation (HFRI), Project Number 2288 (Greece); the Hungarian Academy of Sciences, the New National Excellence Program - \'UNKP, the NKFIH research grants K 131991, K 133046, K 138136, K 143460, K 143477, K 146913, K 146914, K 147048, 2020-2.2.1-ED-2021-00181, and TKP2021-NKTA-64 (Hungary); the Council of Science and Industrial Research, India; ICSC -- National Research Center for High Performance Computing, Big Data and Quantum Computing, funded by the EU NexGeneration program (Italy); the Latvian Council of Science; the Ministry of Education and Science, project no. 2022/WK/14, and the National Science Center, contracts Opus 2021/41/B/ST2/01369 and 2021/43/B/ST2/01552 (Poland); the Funda\c{c}\~ao para a Ci\^encia e a Tecnologia, grant CEECIND/01334/2018 (Portugal); the National Priorities Research Program by Qatar National Research Fund; MCIN/AEI/10.13039/501100011033, ERDF ``a way of making Europe", and the Programa Estatal de Fomento de la Investigaci{\'o}n Cient{\'i}fica y T{\'e}cnica de Excelencia Mar\'{\i}a de Maeztu, grant MDM-2017-0765 and Programa Severo Ochoa del Principado de Asturias (Spain); the Chulalongkorn Academic into Its 2nd Century Project Advancement Project, and the National Science, Research and Innovation Fund via the Program Management Unit for Human Resources \& Institutional Development, Research and Innovation, grant B37G660013 (Thailand); the Kavli Foundation; the Nvidia Corporation; the SuperMicro Corporation; the Welch Foundation, contract C-1845; and the Weston Havens Foundation (USA).  
\end{acknowledgments}
\bibliography{auto_generated}
\cleardoublepage \appendix\section{The CMS Collaboration \label{app:collab}}\begin{sloppypar}\hyphenpenalty=5000\widowpenalty=500\clubpenalty=5000
\cmsinstitute{Yerevan Physics Institute, Yerevan, Armenia}
{\tolerance=6000
A.~Hayrapetyan, A.~Tumasyan\cmsAuthorMark{1}\cmsorcid{0009-0000-0684-6742}
\par}
\cmsinstitute{Institut f\"{u}r Hochenergiephysik, Vienna, Austria}
{\tolerance=6000
W.~Adam\cmsorcid{0000-0001-9099-4341}, J.W.~Andrejkovic, T.~Bergauer\cmsorcid{0000-0002-5786-0293}, S.~Chatterjee\cmsorcid{0000-0003-2660-0349}, K.~Damanakis\cmsorcid{0000-0001-5389-2872}, M.~Dragicevic\cmsorcid{0000-0003-1967-6783}, P.S.~Hussain\cmsorcid{0000-0002-4825-5278}, M.~Jeitler\cmsAuthorMark{2}\cmsorcid{0000-0002-5141-9560}, N.~Krammer\cmsorcid{0000-0002-0548-0985}, A.~Li\cmsorcid{0000-0002-4547-116X}, D.~Liko\cmsorcid{0000-0002-3380-473X}, I.~Mikulec\cmsorcid{0000-0003-0385-2746}, J.~Schieck\cmsAuthorMark{2}\cmsorcid{0000-0002-1058-8093}, R.~Sch\"{o}fbeck\cmsorcid{0000-0002-2332-8784}, D.~Schwarz\cmsorcid{0000-0002-3821-7331}, M.~Sonawane\cmsorcid{0000-0003-0510-7010}, S.~Templ\cmsorcid{0000-0003-3137-5692}, W.~Waltenberger\cmsorcid{0000-0002-6215-7228}, C.-E.~Wulz\cmsAuthorMark{2}\cmsorcid{0000-0001-9226-5812}
% ...
\par}
\vskip\cmsinstskip
\dag:~Deceased\\
$^{1}$Also at Yerevan State University, Yerevan, Armenia\\
$^{2}$Also at TU Wien, Vienna, Austria\\
% ...
\end{sloppypar}
\end{document}